**Title**
Addressing missing context in regulatory variation across primate evolution

**Short Title**
Addressing missing context in primates

**Authors**
Genevieve Housman[1,*], Audrey Arner[2,3], Amy Longtin[2,3], Christian Gagnon[1], Arun Durvasula[5,6], Amanda Lea[2,3,4,*]

**Affiliations**
[1]Department of Primate Behavior and Evolution, Max Planck Institute for Evolutionary Anthropology, Leipzig, Germany
[2]Department of Biological Sciences, Vanderbilt University, Nashville, TN, USA
[3]Evolutionary Studies Initiative, Vanderbilt University, Nashville, TN, USA
[4]Vanderbilt Genetics Institute, Vanderbilt University, Nashville, TN, USA
[5]Department of Population and Public Health Sciences, Keck School of Medicine, University of Southern California, Los Angeles, CA, USA
[6]Center for Genetic Epidemiology, Keck School of Medicine, University of Southern California, Los Angeles, CA, USA
*Corresponding authors

*Correspondence to:*
Email: genevieve_housman@eva.mpg.de
Address: Deutscher Platz 6, 04103 Leipzig, Germany
Phone: +49 0341 3550-224

Email: amanda.j.lea@vanderbilt.edu
Address: VU Station B, Box 35-1634, Nashville, TN 37235
Phone: +1 615 322-2008

**Abstract**

In primates, loci associated with adaptive trait variation often fall in non-coding regions. Understanding the mechanisms linking these regulatory variants to fitness-relevant phenotypes remains challenging, but can be addressed using functional genomic data. However, such data are rarely generated at scale in non-human primates. When they are, only select tissues, cell types, developmental stages, and cellular environments are typically considered, despite appreciation that adaptive variants often exhibit context-dependent effects. In this review, we 1) discuss why context-dependent regulatory loci might be especially evolutionarily relevant in primates, 2) explore challenges and emerging solutions for mapping such context-dependent variation, and 3) discuss the scientific questions these data could address. We argue that filling this gap will provide critical insights into evolutionary processes, human disease, and regulatory adaptation.

**Main Text**

Non-coding genetic variation, which can influence gene regulation rather than protein function, plays a crucial role in shaping adaptive phenotypic variation in humans and other species [1,2]. For example, one of the strongest signatures of positive selection in the human genome occurs near regulatory variants upstream of the *LCT* gene, conferring lactase persistence [3]. Another example is the evolution of human skin pigmentation variation, where light skin color in southern African populations is associated with regulatory variation near *MITF*, *LEF1*, and *TRPS1* [4]. Most efforts to understand adaptively relevant regulatory variants in the primate lineage have been conducted in humans, but incorporating comparative perspectives can provide essential insight. For example, comparative data enable the identification of conserved, essential regulatory elements, as well as genetic variation that is unique to specific taxa, suggestive of lineage-specific importance. A recent analysis of whole genomes from 49 primate species identified a region of accelerated evolution in gibbons involved in limb development, and thus potentially important for this taxa's unique morphology [5]. Similarly, multi-species genomic comparisons have identified regulatory variants at several melanogenesis-related genes in the gray snub-nosed monkey, potentially contributing to the mosaic color patterns of this species [6].

Despite great interest in understanding regulatory variation in humans and our close relatives, identifying the functional role of non-coding variants can be challenging because it is difficult to predict function from sequence alone. Several solutions have emerged to address this problem at genome-scale in humans. For example, consortia such as ENCODE [7], ROADMAP [8], or the Genotype-Tissue Expression project (GTEx) [9] have generated comprehensive functional genomic data to annotate regulatory elements, as well as genetic variants that impact gene regulatory traits across organs, tissues, and cell types (e.g., using transcriptomic, chromatin accessibility, DNA methylation, histone modification, or other functional genomic assays) [10–13]. A second strategy is to use experimental approaches, such as massively parallel reporter assays (MPRAs), to test whether specific sequences are sufficient to causally drive regulatory activity [14–16]. While studies of gene expression, DNA methylation [17,18], chromatin accessibility [19,20], and even MPRAs [21–23] are emerging for non-human primate (NHP) species, large, well-powered datasets (e.g., across many individuals and/or many species) remain rare. Another complicating factor in understanding regulatory variation across primates is that such variants may exhibit context-dependent effects [9,24–26], varying across tissues, cell types, developmental stages, or cellular environments (**Box 1**, **Figure 1**); however, there is a severe lack of large, multi-context functional genomic datasets for NHPs (**Figure 2**). Focusing on this gap, the goal of this review is to 1) discuss why context-dependent regulatory loci might be especially important to characterize if we want to understand adaptive trait evolution in primates, 2) explore logistical and technical challenges, as well as emerging solutions, for mapping context-dependent functional genomic variation in this lineage, and 3) discuss the exciting scientific questions that could be unlocked with these datasets.

**Why is it important to characterize context-dependent regulatory variation across primates?**

Evolutionary theory suggests that context-dependent regulatory mutations may be especially important for complex traits. This is because mutations are only tolerated if they do not incur a net fitness cost to the organism: while mutations with ubiquitous function are likely to generate deleterious effects in some tissues, cell types, developmental states, or environmental conditions, context-dependent mutations can provide targeted, advantageous functions with fewer deleterious, pleiotropic effects [25,27–29]. Consistent with this idea, the GTEx project examined the genetic basis of human gene expression variation across 54 tissues and found 1) that only ~25% of expression quantitative trait loci (eQTLs) have ubiquitous effects on expression (shared across ≥3 tissues), 2) that eQTLs with more tissue-specificity have greater evidence for organism-level impacts on complex traits and diseases, and 3) that eQTLs with more tissue-specificity show stronger signatures of selection [9]. Similar work has expanded this idea across cellular environments: many studies in humans have now identified additional types of context-dependent eQTLs which have different effect sizes at baseline versus after *ex vivo* cellular treatment with pathogens, other molecules that provoke an immune response, drugs, hormones, chemicals, and additional stimuli [30–34]. These studies have consistently shown that such context-dependent eQTLs overlap genome-wide association (GWAS) hits for complex traits and diseases; in many cases, this overlap is stronger than for eQTLs that are constant or "ubiquitous" across cellular conditions [30,32]. Similar to GTEx studies, these types of context-dependent eQTLs are often more strongly enriched for signatures of past adaptation than ubiquitous eQTLs. However, while this work makes a strong argument that variants with context-dependent function may be especially important for evolution and disease, these efforts have yet to be applied widely across broader phylogenetic contexts. We argue that expanding this work to include NHPs is especially important for understanding conservation versus "uniqueness" of human context-dependent regulatory function, as well as evolutionary processes in other primate lineages.

**What are the challenges and solutions for identifying context-dependent effects across NHP species?**

The delay in systematic characterization of context-dependent regulatory variation in NHPs is due, in part, to a variety of ethical and logistic challenges associated with obtaining well-preserved samples, as well as historically having limited numbers and qualities of NHP genome assemblies. Most countries have laws and regulatory bodies (e.g., Institutional Animal Care and Use Committee in the USA) that define and enforce the scope of allowable animal research, and there is general and rightful consensus that the use of NHPs in research should be reduced, refined, and replaced as much as possible [35,36]. Despite these challenges, steady contributions over the past several decades, in combination with exciting technological advances over the past couple years, have drastically improved the genomic data available for NHPs (**Box 2**).

With regards to understanding the adaptive value of regulatory variants, much of this work has focused on characterizing whether putatively interesting regions of primate genomes have regulatory function, regardless of context. Such questions can be answered by inserting candidate genetic material into mouse models (e.g., LacZ transgenic reporter assays) or common human cell lines (e.g., luciferase reporter assays in HEK293T cells) and measuring regulatory outputs [5,6]. While this work generates useful information, it lacks consideration for tissue, cell type, developmental stage, and environmental context. It is also disconnected from NHP primary tissues and their endogenous functions. Fittingly, recent efforts are working to fill these gaps.

**Studying multi-tissue, cell type, and developmental stage effects**

Although collecting primary samples from NHPs is challenging, it is not impossible when appropriate ethical considerations are made. Thus, there have been many calls to better formalize ethical, large-scale collections of comparative functional genomics data from NHP species [37,38], and this work is just now beginning in earnest. For example, like the GTEx project [39], there are now initiatives to collect similar data types from a small number of NHPs – namely macaques [40], marmosets [40], and baboons [41]. In particular, the recently initiated Non-Human Primate Developmental Genotype-Tissue Expression (NHP dGTEx) project [40] plans to systematically collect dozens of tissues from 126 macaques at the Oregon National Primate Research Center and 72 marmosets from the Massachusetts Institute of Technology at different ages, characterize gene expression and chromatin accessibility patterns at bulk and single-cell resolutions, and test how these molecular phenotypes relate to underlying DNA sequencing variation. In this project, context-dependent regulatory variation will be examined in relation to tissue, cell type, and developmental stage, consistent with recent studies examining regulatory patterns in NHP tissues [42–46]. Nevertheless, there are some specific contexts (like skeletal tissues) that will be excluded from this project, as well as many other contexts, such as external environmental exposures, that will not be considered.

As an alternative to primary tissue sampling, many researchers have worked towards establishing, expanding, and conducting experiments with cell culture resources. For example, lymphoblastoid cell lines (LCLs) from large panels of humans, made possible through the 1000 Genomes Project Consortium [47], have generated substantial functional genomics data. While generating LCLs (B-cells immortalized using the Epstein Barr or related virus) is one of the oldest cell line establishment methods – beginning about six decades ago [48–51] – similar large-scale studies in NHP LCLs have not yet been conducted; although, a small panel of ape LCLs has recently been used to more deeply characterize epigenetic landscapes across species [20]. Currently, the most important cell culture technologies for functional genomics research are induced pluripotent stem cells (iPSCs). These cells are reprogrammed from adult cell types without altering the underlying genotypes, allowing them to self-renew and have the potential to differentiate into any cell type. Current nonhuman iPSCs, primarily derived from apes but also from other NHPs [52–56] and even more evolutionarily diverse taxa (e.g., Stem Cell Zoo [57]), have been generated from primary cells (e.g., blood or fibroblasts) that were obtained through "invasive" sampling of tissue (we note that even samples obtained from peripheral tissues or opportunistically as byproducts from necessary medical procedures are still considered invasively collected). Only recently have researchers attempted – with moderate success – to establish NHP iPSCs from completely non-invasive tissues such as urine [56].

While gene expression is the molecular phenotype most often characterized in primate iPSCs, several other molecular phenotypes have been studied including transposable element regulation [58], chromatin accessibility [59], and chromatin folding [60]. Researchers have also developed interesting ways to study the factors contributing to gene expression variation. For instance, recent studies that fuse human and chimpanzee iPSCs to form allotetraploid composite cell lines [61] have allowed researchers to disentangle *cis*- and *trans*- effects in closely related primate species [62,63]. Additionally, gene editing technologies are increasingly being incorporated into iPSC work. As an example, a recent study performed genome-wide CRISPR interference screens in human and chimpanzee iPSCs and identified 75 genes that have species-specific effects on cell cycle and cellular proliferation [64].

However, perhaps the most exciting aspect of iPSCs is their ability to differentiate into other cell types. Primate iPSCs have undergone directed differentiation into a variety of cell types – including cardiomyocytes [65], endoderm cells [66], brain organoids [67], and skeletal cells [68] – to examine steady-state gene expression differences across species. Excitingly, there are now proposals to begin using heterogeneous differentiating cultures, such as spontaneously and asynchronously differentiating iPSC-derived embryoid bodies, alongside single-cell sequencing to capture regulatory patterns from many cell types at different stages of differentiation simultaneously [69–71]. Although this system has some limitations regarding the range of possible cell types, it may greatly aid GTEx-like efforts while not requiring further invasive tissue sampling.

**Studying multi-environmental context effects**

Another exciting feature of NHP cell culture systems is that they can be exposed to a variety of controlled, environmental perturbations *in vitro*. This design offers a clear path toward identifying context-dependent regulatory variation as compared to NHP *in vivo* research, which requires immense efforts to ensure that living organisms are exposed to diverse and controlled environments. Such a design is not impossible; for example, recent work exposed almost one-hundred baboons to a low-cholesterol, low-fat (LCLF) diet for 2 years and then a high-cholesterol, high-fat (HCHF) diet for 2 years. At the end of each diet period, tissue biopsies (adipose, muscle, and liver) were obtained from each individual, and gene expression patterns were measured. By incorporating known genotypes, researchers were also able to map over ten-thousand eQTLs, a subset of which were diet-responsive eQTLs enriched in human GWAS hits for metabolic diseases. Of particular interest, researchers found that diet-responsive eQTLs, as compared to steady-state eQTLs, have genomic locations and connectivity features that are more similar to GWAS hits. This important finding adds to an ongoing discussion [25,72] and provides evidence supporting the idea that response eQTLs are particularly relevant for traits and diseases.

As noted above, given the challenges of *in vivo* studies, most research on environmentally-dependent genetic effects has focused on *in vitro* designs. While these studies are exciting and rapidly expanding in NHPs, they still often do not have large enough sample sizes to detect eQTLs (or other molecular QTLs). For instance, a recent study differentiated human and chimpanzee iPSCs into cardiomyocytes, and exposed these cells to control, hypoxic, and re-oxygenation conditions to examine how dynamic gene expression responses to cardiovascular disease-related environmental perturbations vary across species [73]. Although the sample size (8 humans and 7 chimpanzees) was too small for eQTL testing, the researchers were able to identify hundreds of species-specific regulatory responses across contexts, as well as conserved regulatory responses. In parallel human research, perturbation screen studies are becoming larger and are more commonly characterizing regulatory responses to multiple environmental exposures simultaneously [30,33,34,74]. To add complexity, there are now even efforts to combine perturbation screens with heterogeneous differentiating cultures to examine regulatory response across all of these contexts at the same time [75]. Currently, these advances are restricted to humans, and more work is needed to advance similar studies in NHP cell lines.

Overall, while these functional genomic efforts are increasing the resolution of regulatory landscapes known in NHPs, continued work using NHP primary tissues is likely to remain limited. Thus, there is continued interest in using other suitable models, and NHP cell culture systems will likely continue to fill this gap. Although the potential of these systems is substantial, there are also several caveats. First, *in vitro* cells do not necessarily perfectly replicate *in vivo* cells. This is because *in vitro* and *in vivo* environmental conditions can differ, as well as cell differentiation protocols and developmental processes. Further, a given cell type is not necessarily identical across all ontogenetic stages or regions of the body, so we should not assume that a cell type of interest is completely uniform across the body. Lastly, given the simplicity of cell culture systems as compared to organisms, we may miss some of the complex pathways that connect stimuli and response at the organismal level [76]. Nevertheless, continued efforts to establish new cell lines from additional individuals and species, as well as using previously established cell lines in new ways, will allow us to explore more regulatory phenotypes, more cell types, and more environmental contexts.

**What are the scientific gains from filling these gaps?**

First, expanding our understanding and annotation of context-dependent regulatory function in NHPs is important for identifying the substrates of adaptive evolution in these species. In particular, many recent NHP genome sequencing studies have identified potentially important loci for adaptive trait variation through statistical scans for selection [5,6,77–79]. When implicated loci fall outside of protein-coding regions, it is assumed that they impact traits through changes in gene regulation. However, confirming such mechanisms of action requires actually measuring regulatory patterns (e.g., gene expression, DNA methylation, chromatin accessibility). By filling in our knowledge gaps of functional response to perturbation, we can clarify the mechanisms driving these signals of positive selection and add to conversations about the role of ubiquitous versus context-dependent regulatory variation in adaptive evolution.

Second, studying functional genomic variation across NHPs, including through a context-dependent lens, can help us understand the causes of complex human diseases. For example, recent studies have shown that non-coding sequences conserved across mammals and especially across primates are highly enriched (6- and 10-fold, respectively) for disease heritability, dramatically more so than eQTL or regulatory element annotations commonly used to prioritize non-coding SNPs [80,81]. These results highlight the power of evolutionary conservation for prioritizing disease-relevant SNPs, but without comprehensive functional data across the evolutionary tree, we 1) lack a mechanistic understanding of why these annotations are so successful and 2) we may be missing an opportunity to combine functional genomic and evolutionary annotations to further improve prioritization. Studying conservation of functional genomic variation can thus augment studies of conservation of sequence and help us understand why conserved regions of the genome tend to be enriched for disease heritability.

Third, dissecting the genetic architecture of context-dependent effects across species will help to reveal if certain taxa are especially environmentally perturbable or if responses are generally conserved. For example, humans have displayed high sensitivity to certain immune-related exposures as compared to more distantly related NHPs [82]; however, until recently, it has been unclear whether such context-dependent effects are definitively unique to humans. In fact, humans and nonhuman apes appear to share the capacity for strong and sensitive early responses to infection, which was only revealed by incorporating additional species (humans, chimpanzee, macaques, and baboons) into a perturbation study design [83]. Continuing to increase both the phylogenetic distribution of primate species, as well as the number and types of perturbations, will help resolve the degree to which responses are evolutionarily conserved or divergent across a range of environmental contexts with relevance to differing traits.

Fourth, mapping context-specific regulatory variation across primates will open new avenues of research for methods development and theory. For example, current approaches to studying how context influences adaptation often rely on separate statistical tests for context-specificity and adaptation [31–33]. Because each test can produce false negatives, performing them independently may underestimate their overlap. Future work could increase statistical power by jointly modeling both context-dependence and adaptation within an extended quantitative genetic framework [84]. In particular, quantitative genetic theory commonly addresses context-specificity through the lens of phenotypic plasticity, which is a genotype's ability to produce different phenotypes in different environments [85]. Some studies have extended standard quantitative genetic models to capture genetic variation for plastic responses [84,86,87], while others have focused on regulatory genes that control this plasticity [88]. Further work is needed to interpret context-dependent molecular QTL in a quantitative genetic framework. Recent findings suggest that amplification of genetic effects may be a key driver of genotype-by-environment interactions [89,90], but these insights have yet to be incorporated into current theoretical models. Incorporating these new perspectives holds promise for refining quantitative genetic theory and enhancing our predictions of evolutionary responses to changing environments across diverse species.

**Conclusions**

This review argues the importance of characterizing genetic effects on regulatory element function across NHPs, especially across contexts (including life stages, tissue/cell types, and environmental contexts). Doing so will provide key insights into primate evolution and adaptation, as well as why evolutionarily conserved regions are enriched for human diseases, and drive new methods, theory, and other developments across diverse fields of biology. We highlighted current work that is beginning to realize this vision using primary samples collected across NHP tissues and developmental time points, controlled whole organism environmental manipulations, and myriad *in vitro* and cell culture approaches. While exciting, several critical gaps remain. First, while powerful, essentially none of the study designs we focused on capture natural environmental variation, thus making them somewhat removed from the social, nutritional, and ecological contexts in which NHP species evolve. Biobanking during opportunistic events in natural populations provides one path forward [91,92], but connecting natural variation to genome function will likely always be a challenge (as it is in essentially all organisms). Second, while most of the studies we discuss focus on gene expression as their molecular phenotype of interest, recent work in humans has highlighted the importance of simultaneously characterizing the context-dependent genetic architecture of the epigenome [93]. We thus look forward to more multi-omic work across NHP species, especially with sample sizes appropriate for mapping genetic architecture. Finally, human studies of context-dependent genetic effects at the molecular level currently benefit from vast annotation databases to link mechanism and organism-level traits: in other words, given a set of context-dependent QTL, it is feasible to overlap them with catalogs of population and evolutionary genomic variation [94], GWAS hits [95], transcription factor binding events [8], and more. Moving forward, we hope the resources for interpreting the mechanistic and phenotypic relevance of candidate loci in NHPs grows hand in hand with functional genomic datasets. Together, these efforts will catapult our understanding of our close evolutionary relatives, as well as ourselves.

**Funding**

This work was supported by the Kinship Foundation (Searle Scholars Program), the National Institutes of Health (R35GM147267), the National Science Foundation (Graduate Research Fellowship Program 2444112), and the Vanderbilt Evolutionary Studies Initiative.

**Disclosures**

The authors declare no conflicts of interest.

**Figures**

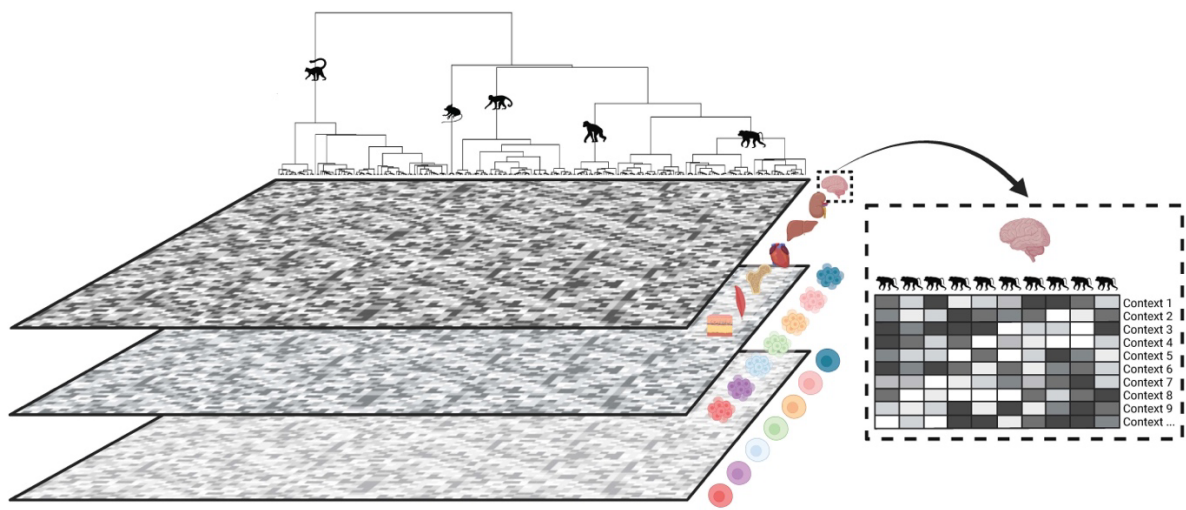

**Figure 1. Dimensions of context-dependent regulatory variation to be characterized across primates.**
Depiction of the various functional genomic datasets that could be collected across tissues, cell types, developmental stages, and cellular environments for diverse primate species to identify context-dependent regulatory loci. Heatmaps represent the possibility of measuring molecular phenotypes (e.g., gene expression) from different contexts across different individuals, subspecies, and species within the primate order (note the heatmap values themselves are not biologically based nor do they represent biological hypotheses). In the stacked heatmaps, each level represents variation across developmental stages, ranging from undifferentiated, pluripotent cells (bottom), to differentiating and maturing cells (middle), and lastly fully differentiated cells (top) across a variety of different tissues. In the right panel with the 2-dimensional heatmap, each row represents variation across different cellular environments for a given cell type, in a given tissue, for a given set of primate individuals. The primate phylogenetic tree was generated using data from the 10kTrees website Version 3 [96], and portions of this figure were created with BioRender.com.

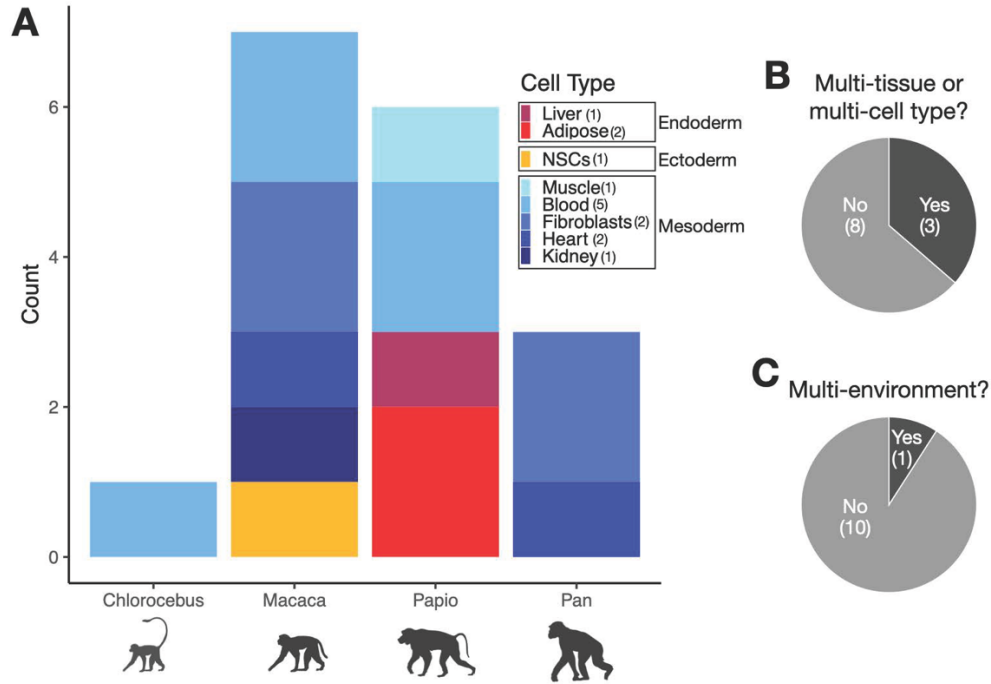

**Figure 2. Existing expression QTL studies conducted in NHPs.**
Summary of findings from a literature review of existing expression QTL studies in NHPs (see **Supplemental Text**). **(A)** Histogram displaying the number of studies that include samples from a given genus, with colors denoting the tissue or cell type examined in each study. Some studies include more than one species and/or more than one tissue or cell type. In the figure legend, the numbers listed next to each tissue or cell type represent the number of studies that include such samples. **(B)** Pie chart depicting the number of studies that contained multi-tissue or multi-cell type analyses ('Yes') or were conducted in only one tissue or cell type ('No'). **(C)** Pie chart depicting the number of studies that contained imposed, experimental environments ('Yes') or were conducted at baseline ('No'). Silhouette images were created with BioRender.com.

**Boxes**

**Box 1. Defining context-dependent effects.**
There are many ways that 'context-dependent' can be defined within the field of functional genomics. Broadly, we define 'context-dependent' research as investigating variations in tissue, cell type, developmental stage, or cellular environment. Here, we define each type of context-dependent effect and provide experimental designs that can identify such effects. We note that this is not an exhaustive list but rather the methods most discussed in this paper.

Multi-tissue datasets can be considered context-dependent given the distinct gene expression and epigenomic patterns of various tissue types [9,97–101]. Thus, each tissue type can be considered its own context across which genetic effects are compared..

Similar to multi-tissue analyses, single-cell studies (e.g., scRNA-seq, snATAC-seq, snRNA-seq, scDNAm, etc.) in heterogeneous tissues can also be considered context-dependent research. Through such methodologies, it is now clear that individual cell types, even within tissues, produce distinct genomic profiles [102–104] indicating the context-specificity of individual cell types.

Varying developmental stages can also create diverse cellular environments in which the same genome can be studied. Broadly, research comparing differentiation stages in cell culture systems, as well as work in primary tissues, has found developmental stage specificity of gene expression patterns and certain epigenetic markers [105–109]; thus, marking developmental stages as another 'context' through which the genome can be studied.

Finally, environmental exposures, whether imposed naturally or experimentally, are another aspect of context-dependence. Examples include *in vitro* and *in vivo* studies (e.g., [83] and [24]). Here, study subjects, or their biological materials, are exposed to a given environmental perturbation which is then compared to other conditions or a control. From these experiments, researchers are able to compare the genetic architecture of gene regulation at baseline and during perturbations.

**Box 2. The breadth and quality of NHP reference genomes.**
To characterize regulatory landscapes and their genetic basis, it is essential to have high quality genome assemblies and well-characterized patterns of genetic variation. Although the first NHP reference genome (chimpanzee [110]) was completed two decades ago – and only a few years after the human reference genome [111] – NHP genome assemblies and annotations have lagged as compared to those for humans and other model organisms. Additionally, until recently, only a subset of the over 500 NHP species even had assemblies, which has limited functional genomics research in these taxa. However, in 2023 the phylogenetic breadth of available NHP genome assemblies increased to almost half of the entire order (n=233 species) after a collection of papers reported newly generated assemblies for 211 primate species [112]. Perhaps more importantly, there are now telomere-to-telomere (T2T) reference genomes being generated for several NHPs, including apes [113,114], macaques [115], and more still in-progress [116]. Using long-read sequencing – as was done for the human T2T genome a couple years earlier [117] – these assemblies are gapless, containing information about highly repetitive and complex regions of the genome that are unresolved in more traditional reference genomes. These genomic resources provide new insights into DNA sequence and structural variation, and, when combined with other whole-genome sequencing collections are helping to clarify deep evolutionary divergences [77,118], as well as more recent events of admixture from ghost lineages [119,120] and hybridization of ancestral [6,121] and extant [122,123] species. Altogether, these genomic data are essential stepping-stones for further functional genomics research.

**Supplementary File Descriptions**

**Supplemental_Text.docx**
Document describing the methods and results of a literature review that was conducted to generate **Figure 2**.

# Supplemental materials for: Addressing missing context in regulatory variation across primate evolution

Genevieve Housman, Audrey Arner, Amy Longtin, Christian Gagnon, Arun Durvasula, Amanda Lea

**Supplemental Methods**

On February 19th, 2025, we conducted a brief literature search using NCBI's PubMed Advanced Search Builder [1] and the Britanica list of primates [2]. For each search, we entered a primate group or species as one search term within all fields and 'eQTL' as a second search term within all fields (e.g. '(baboon) AND (eQTL)'). This was repeated for all primates listed in the Britanica list of primates. The search results were exported as .csv files to obtain metadata. These results produced 24 citations, 11 of which were unique and relevant to the search criteria (see **Table S1**). These papers were then manually scanned to extract species, tissue and cell type, experimental design, type of genomic data, and sample size information.

| DOI | Publication Year | Multi-tissue or multi-cell type | Cell type(s) | Multi-environment | Environment(s) | eQTL | Multi-species | Species | Sample Size |
|---|---|---|---|---|---|---|---|---|---|
| 10.7554/eLife.04729 | 2015 | no | blood | no | NA | yes | no | baboon | 63 |
| 10.1016/j.xgen.2024.100509 | 2024 | yes | liver; muscle; adipose | yes | low-cholestoral; high-cholesterol | yes | no | baboon | 99 |
| 10.1038/hdy.2008.28 | 2008 | yes | adipose; blood | no | NA | yes | no | baboon | 404 |
| 10.1093/hmg/dds160 | 2012 | no | blood | no | NA | yes | no | vervet monkey | 626 |
| 10.1186/s12864-017-3531-y | 2017 | yes | blood; heart; kidney | no | NA | yes | no | macaque | 46 |
| 10.1002/ajmg.b.32835 | 2021 | no | neural stem cells | no | NA | human | no | macaque | NA |
| 10.1534/genetics.118.301028 | 2018 | no | fibroblasts | no | NA | human | yes | human; chimpanzee; macaque | NA |
| 10.1111/mec.17576 | 2024 | no | blood | no | NA | meQTL | no | macaque | 573 |
| 10.7554/eLife.59929 | 2020 | no | heart | no | NA | yes | yes | human; chimpanzee | 78 |
| 10.1186/s12862-022-02020-x | 2022 | no | NA | no | NA | human | yes | human; chimpanzee | 69 |
| 10.3389/fgene.2019.00152 | 2019 | no | fibroblasts | no | NA | human | yes | human | NA |

**Table S1. Literature review results.** Unique results from literature review described in the Supplemental Methods. Data from this table was used to generate **Figure 2**.

# References

1. **Advanced Search Results**. *PubMed* 2025,

2. The Editors of Encyclopedia Britannica: **list of primates**. *Encyclopedia Britannica* 2016,